# The lunar eclipse and the dawn of astrophysics in van Gogh's masterpieces

Edvige Corbelli
*INAF-Osservatorio Astrofisico di Arcetri, L.E. Fermi 5, 50125 Firenze, ITALY*
edvige.corbelli@inaf.it

**Abstract**

When visiting the Museum of Modern Art's special exhibit "van Gogh and the Colors of the Night (New York, USA, 2008)" I was impressed by the artist's effort to paint and write about landscapes and life scenes at night. At the same time a clear contrast emerged between the colored Starry Night paint and the darkness or twilight in his earlier production. Since then, while revising van Gogh's artistic production and reading his letters I have come to the conclusion that his work was not only driven by the poetry of the night but also by his awareness of the dawn of astrophysics in the 19th century. A change in the way the artist conceived the sky after his stay in Paris is evident in his later work, with the skies becoming more dynamic and rich in colors. In this paper I am presenting new elements in van Gogh's masterpieces that refer to astronomical events visible from France in 1889 and 1890, such as a lunar eclipse and planetary conjunctions. I will discuss these elements and analyze what might have inspired the depiction on canvas of stars with different structures and colors as well as the large neutral central swirl in the Starry Night. We are left with the impression that the artist had a profound attention to celestial phenomena and some knowledge of the new wonders and discoveries of astrophysics in 19th century.

**1. Introduction**

There is nothing more compelling than the stars in a night sky to fulfill the human mind's need to go and explore, just as there is nothing more inspiring than a starry night for an artist to visualize the invisible. The relationship between these two different sides of cultural heritage, arts and astronomy is long lasting and it is enhanced by the use of images from both disciplines to communicate. Before the end of XIX century, when the first photographs of celestial objects became available, astronomers used hand drawings to convert telescopic visions into printed images. This exercise required the ability to transfer onto a piece of paper the relative position of point-like objects, like stars and planets, but also the shape and morphology of more diffuse objects like nebulae. Artists, like Étienne Léopold Trouvelot for example, were engaged by astronomers to use telescopes and produce detailed astronomical illustrations. However, in general artwork may not be illustrative of the real world or a faithful reproduction of what eye can see. Artists can distort the vision of a material world to transfer scientific discoveries into their works. They can use imagination, and dare to go beyond what we see and know. The surrealist painter Salvador Dalí, for example, created works with distorted clocks and extra dimensions of space, perhaps inspired by modern concepts from general relativity that gave new insights into space and time.

In this framework we analyze some of the world-renowned artistic paintings by Vincent van Gogh related to night skies. Evenings and nights have been constant sources of inspiration to many artists as to Vincent van Gogh, who painted masterpieces with scenes and landscapes at night and at twilight, beginning with his early work of The Potato Eaters. There have been many papers and artistic exhibitions focusing on the feelings that have driven the recurrent night scenes in van Gogh artistic production, such as religious, philosophical, and poetic thoughts (see for example the exhibit *van Gogh and the colors of the night,* van Heugten et al. 2008). Here we would like to focus on a new element that might have played a fundamental and key role in driving some of his latest artistic production: the artist awareness of celestial phenomena and his knowledge of scientific discoveries at the heart of astrophysics, a new discipline that was emerging in the XIX century, aimed at understanding the physical nature of stars and other celestial bodies.

A few art historians and professional astronomers in the past have already given some hint that astronomy might have influenced the life and artistic production of van Gogh, given the artist effort to display in his artistic canvas specific observations of the night sky. The historian Boime (1984) asserted that although van Gogh in his letters never mentions astronomy or Camille Flammarion, author of popular astronomy books, he must have been aware of illustrations of the spiral nebulae and might have represented one of these, at the center of his Starry Night. Whitney (1986) agrees on this possibility but suggested that the swirls in the sky could also represent the mistral wind in Provence. Whitney believed in van Gogh attempt to reproduce what he observed in the night sky and estimated some painting dating by comparing the positions of astronomical objects in van Gogh's work with estimated real locations in the sky. Following Whitney, Olson et al. (2003) reported the results of an expedition in southern France dating van Gogh's Moonrise by comparing the moon position in the painting with the observed position, after corrections done for lunar cycles with the help of a computer.

Here we shall point out new elements present van Gogh's artistic paintings that unveil and underline the artist constant attention to celestial events as well as his profound knowledge of astrophysics progresses in XIX century. After brief remarks to the historical and geographical context of the artist's life, we analyze well-known paintings, illustrative of the tight connection between art and science in van Gogh's production.

## 2. Few facts on van Gogh's life and its historical context

By reading books and journal articles dedicated to van Gogh life and artistic production (e.g. Stone 1934, Scherjon and de Gruyter1937, Hulsker 1980, Callow 1990, Naifeh and Smith 2012) as well as the letters he wrote to his brother Theo, to family members and friends (for these we refer to the English translation and collection quoted as V. van Gogh 2009[1]), one realizes that the artist has had a very humble personality and at the same time has been culturally very rich. Having worked as a book seller at Dordrecht in 1877 for a few months, he had the opportunity to know and read several books and publications and to translate passages from the Bible into English, French and German. His relation with arts started in 1869 when he obtained a position in the Hague branch of the art dealers Goupil & Compagnie. After a training period he moved to London and later to Paris to work for other Goupil's branches until 1876. We underline this to remark that at age 23 he had already travelled across Europe (Ramsgate, Isleworth, Dordrecht, Amsterdam, Brussel, Drenthe, Nuenen, Antwerp) and he continued to do so for the next decade.

We would like to recall here that Vincent van Gogh spent about two years in Paris, between 1886 and 1888, before moving to Arles in southern France. In Paris he first lived with his brother Theo at 54 Rue Lepic in Montmartre where he was introduced to the work of the impressionists. The area was an artist's quarter and a place with cafés and circles such as Le Chat Noir. Here van Gogh met artists and writers. He became friends with Émile Bernard, Paul Signac and Paul Gauguin. This last artist illustrated one of Cros's poem. Charles Cros was a major figure in Le Chat Noir and one of Flammarion's closest friend, connected to Impressionist painters. Boime (1984) describes the great influence that Flammarion could have had on van Gogh's thoughts and in particular in relating the great space in the cosmo with religion and with the possibility of travelling in space. Flammarion's name is never explicitly mentioned in van Gogh's letters and we don't know if the two have ever met or if van Gogh knew about Flammarion's ideas indirectly through Charles Cros. Astronomy was surely becoming widespread thanks to Flammarion's popular books (such as Astronomie populaire, Les etoiles, Les pluralite des mondes habites) and to the reproduction of nebulae in magazines (such as L'Astronomie, L'Illustration, Harper's Weekly etc.). The

---

[1] URL https://www.vangoghletters.org/vg/letters.html

impact of Flammarion's connection between astronomy, immortality and life can be read in van Gogh writing:

> "*For my own part, I declare I know nothing whatever about it, but looking at the stars always makes me dream, as simply as I dream over the black dots representing towns and villages on a map. Why, I ask myself, shouldn't the shining dots of the sky be accessible as the black dots on the map of France ? Just as we take the train to get to Tarascon or Rouen, we take death to reach a star.*" (1888, Letter to Theo 638).

van Gogh was excited about maps in general and he used to draw maps as gifts for his friends and family. The International Astronomical Congress which took place in April 1887 at Paris Observatory had as main resolution the use of photographic techniques to create a chart of the heavens (cf. Flammarion 1907). Van Gogh disliked photography. Referring to portraits he writes to his sister:

> *"I myself still find photographs frightful and don't like to have any, especially not of people whom I know and love.*
> *These portraits, first, are faded more quickly than we ourselves, while the painted portrait remains for many generations. Besides, a painted portrait is a thing of feeling made with love or respect for the being represented."* (1889, Letter to Willemien 804).

His attempt to paint stars in the sky with care for their colors, position, and luminosity, as we will see in this paper, might also be an attempt to show that artists can do a better job than photographs. Van Gogh is not a scientist and he is very conscious of his limits, being very humble, he hardly writes explicitly about his desire to do so and about astronomy in general. However there are several elements emerging from the analysis presented in this paper that underline his attention to astronomical events and the likely awareness of new astrophysical concepts emerging at the end of XIX century.

## 3. Moonrise: framing the 1889 lunar eclipse

The painting Evening landscape at moonrise (hereafter referred to as Moonrise) shows wheat stacks in a field enclosed by a mountain ridge in the twilight sky with a prominent orange disk partly hidden behind the mountains (see Figure 1). Van Gogh often looked at the sky from his window on the east side of the Saint Paul monastery in Saint-Rémy de Provence, as he mentions to his brother:

> *"Through the iron-barred window I can make out a square of wheat in an enclosure, a perspective in the manner of Van Goyen, above which in the morning I see the sun rise in its glory. With this — as there are more than 30 empty rooms — I have another room in which to work.*" (1889, Letter to Theo, 776 )

From this it is clear that his window faces east while it is not clear what the orientation of the additional working room is. Jacob-Baart de la Faille, a Dutch art historian and writer, compiled the first critical catalogue of van Gogh's works (de la Faille 1928) and referred to this work (F735) as Sunset. A catalog by Scherjon and de Gruyter (1937) changed the title of the work to identify the scene as Rising Moon, in agreement with the field towards east that can be seen from his window. In 1939 de la Faille in the revised and updated catalog corrected accordingly the painting title.

Dating the Moonrise work has been a long and debated issue. Van Gogh mentions this work in a letter to Theo now dated as Sunday 14 or Monday 15 July 1889:

> *"I have one in progress of a moonrise over the same field as the sketch in the letter to Gauguin, but with wheat stacks replacing the wheat. It is dull yellow ochre and violet. Anyway, you will see it sometime soon." "* (1889, Letter to Theo 789)

The Scherjon and de Gruyter catalogue (1937) placed this canvas in August or September 1889 but the later edition of de la Faille catalogue (1970), published by a team of experts after the author's death, specifies July 6 1889 as the date of the moonrise painting. This confusion arose because apparently the date is missing on the letter where the moonrise work is described. Hulsker (1980), working on van Gogh correspondence, suggested that the letter was written much earlier than August or September, as previously stated, and moved the Letter to July 6. To narrow the uncertainties, an expedition by Olson et al. in 2002 (Olson et al. 2003) journeyed to Saint-Remy de Provence, France, observing the moon rising for several days. Using astronomical calculations and deductive reasoning, they concluded that van Gogh's painting Moonrise depicted at full moon on 9:08 p.m. local mean time on July 13, 1889.

I have always been puzzled by the reddish color of the moon in that painting which appears on one side only, and which has been one of the unusual elements which generated confusion for decades between the moon and sun in that canvas. I therefore searched the NASA Eclipse Web Site for Lunar Eclipses between 1801 and 1900[2] and found that the lunar eclipse number 09393 happening on July 12th 1898 was visible from southern France. This was a partial lunar eclipse (Saros 137) of a duration 142 minutes and an umbral magnitude of 0.48. Using the application STELLARIUM[3] it can be easily seen that in Saint-Remy-de-Provence the penumbra phase of this Lunar Eclipse started soon after the moon rose above the horizon in the SE direction. The umbra shadow starts over the moon's surface when the moon's elevation was 3.3º and 127° degree in azimuth, around 20:05 local time[4]. Due to the cliff, it is likely that the moon and its eclipse became visible from van Gogh room a few minutes later, as the moon rose above the Alpine ridge. The day after, on July 13 1889, the moon crossed the cliff, at 126° in azimuth and 4.5° in altitude, as painted in the canvas, at about 21:08 according to the calculation of Olson et al. (2003). We confirm this conclusion of Olson et al (2003) with the use of STELLARIUM. The day before, the umbra phase of the lunar eclipse in Saint-Rémy-de-Provence was visible until 22:25 and had its maximum around the 21:12 local time, when the moon was about 12° above the horizon and 140° in azimuth. Since the cliff is higher in altitude towards the south (see Figure 1), van Gogh presumably saw the moonrise when eclipse had already started, and he has been able to watch the eclipse throughout its full duration. The eclipse phenomena, with the color of the moon becoming reddish, must have impressed van Gogh who was a careful and frequent observer of the sky. We don't know if he was aware that a lunar eclipse was about to happen on July 12 1889, but he was aware of the existence of lunar eclipses, although he only had a vague idea of the phenomenon. We report a sentence referring to a canvas (Road with Cypress and a Star) painted when the moon was crescent, in the absence of a lunar eclipse (see also Section 6):

*"I still have a cypress with a star from down there, a last attempt - a night sky with a moon without radiance, the slender crescent barely emerging from the opaque shadow cast by the earth."* (1890, Letter to Paul Gauguin RM23)

Inspired by the lunar eclipse, it is likely that Vincent van Gogh started the painting Moonrise the day after. He had clearly in mind the particular color of the moon that he observed on July 12, 1889, but he painted the moon as he saw it on the sky on July 13, around the same time of the night when the maximum umbral eclipse magnitude was visible from Saint-Rémy-de-Provence. According to STELLARIUM the reddish side of the moon was located towards east first, then moved to the north. The artist instead seems to indicate as reddish the south-west part of the moon, emerging from the cliff, as it can be seen in Figure 1. It is possible that he did not remember this detail the following day. However, if he was aware of the fact that an eclipse occurs when the Earth's shadow covers our satellite, he might have by purpose placed the reddish part of the moon behind the cliff, to mimic a shadow, as artificially created by the sunlight from the opposite direction.

---

[2] https://eclipse.gsfc.nasa.gov/LEcat5/LE1801-1900.html

[3] STELLARIUM is an astronomy software that mimics a planetarium URL
https://stellarium.org/

[4] in 1889, as pointed out by Olson, Saint-Remy local time was 19 minutes later than Universal time and we shall use this time in the rest of this Section

During the XIX century, the German astronomer Friedrich Bessel developed a method still in use to facilitate the calculation of local circumstances and conditions of visibility of a solar eclipse. All these developments were mainly possible due to ever-improving knowledge of the distance between the Earth-Moon and Earth-Sun since the XVII century. Therefore we cannot exclude that van Gogh was aware of the lunar eclipse before this occurred. He might have started the canvas the same night he observed the phenomenon although we feel this is unlikely because of the mismatch between the moon position in the canvas and the real one. Eclipses have always been associated with legends, myths and symbols which constitute a rich source of inspiration in different cultures and epochs. On 29 May 1453, a rising full moon was eclipsed over Constantinople, then under siege by the Turkish army. It is reported that this created such a dip in morale that in a few days Constantinople was defeated, leading to the end of the Eastern Roman Empire after 1130 years. It seems normal that at the time of an eclipse one reflects further about the relation of man to Earth, the Moon, the Sun, the cosmos. For these reasons van Gogh might have been particularly inclined and driven to immortalize this unique event.

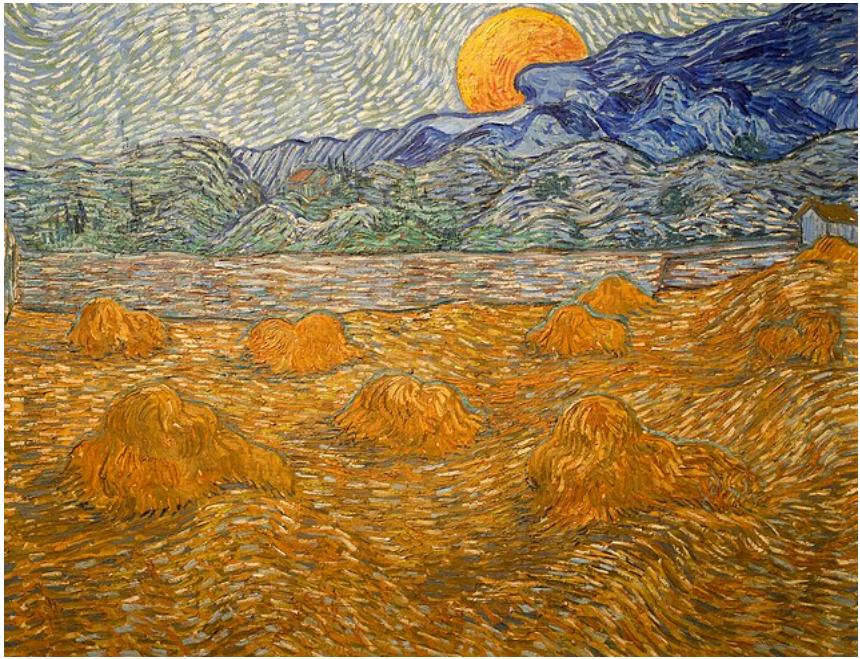

*Figure 1:* **Evening landscape at moonrise**, (V. van Gogh**,** Saint-Rémy de Provence, July 1889). Exhibited in the Kröller-Müller Museum, Otterlo. With the use of STELLARIUM we have discovered that this masterpiece reproduces a rising moon during a lunar eclipse visible in southern France on July 12 1889. This explains the reddish color of part of the moon (The image is under Creative Commons Attribution-Share Alike 4.0 International license).

### 4. Starry nights and the colors of the stars

Van Gogh arrived in Arles in February 1888 and made a number of paintings of the city, communicating at the same time through his letters his strong will to paint the night sky. In April he wrote to his brother Theo: "...*I need a starry night with cypresses or maybe above a field of ripe wheat, there are some really beautiful nights here...*" (1888, Letter to Theo, 594). And to the painter Emile Bernard in the same period he wrote:

> "*A starry sky, for example, well — it's a thing that I'd like to try to do, just as in the daytime I'll try to paint a green meadow studded with dandelions*." (1888, Letter to Emile Bernard 596)

In June once again, he confided to Emile Bernard: "...*But when will I do the starry sky, then, that painting that's always on my mind?..*" (1888, Letter to Emile Bernard, 628). And between Sunday, 9 September and Friday, 14 September 1888 he wrote to his sister:

> "*I definitely want to paint a starry sky now. It often seems to me that the night is even more richly coloured than the day, coloured in the most intense violets, blues and greens. If you look carefully you'll see that some stars are lemony, others have a pink, green, forget-me-not blue glow. And without labouring the point, it's clear that to paint a starry sky it's not nearly enough to put white spots on blue-black.*" (1888, Letter to Willemien 678)

Starry Night over the Rhone, shown in Figure 2, was one of three paintings made during September 1888 that incorporate the night sky and stars (the other two being Cafe Terrace at Night and a portrait of his friend Eugene Boch). The artist walking near home, between the banks of the Rhone river in the city of Arles, discovered a suitable point to represent the starry night. This night scene was prompted by a genuinely moving experience of the endless darkness, an experience van Gogh describes in a letter: "*Once I went for a walk along the deserted shore at night. It was not cheerful, it was not sad - it was beautiful.*" (1890, Letter to Theo, 619). And to his friend Eugène Boch he wrote:

> "*I am working [...] on a study of the Rhone, of the city illuminated by gas lamps reflected in the blue river. Above, the starry sky with the Big Dipper, a glimmer of pink and green on the cobalt blue field of the starry sky, where the lights of the city and its cruel reflections are red gold and bronze green...*" (1888, Letter to Eugène Boch, 693).

In the Starry Night over the Rhone the artist paints stars that form the Big Dipper as in the Ursa Major constellation. We recall that the Ursa Major is the first constellation introduced by Flammarion in the section The Stars and the Sidereal Universe of Popular Astronomy (Flammarion 1907). For this canvas there has also been a study for its dating: the Italian astronomer Masi and Basso[5] carried out a study which allowed them to reconstruct the time and date of Starry Night over the Rhone. They estimated the execution of the painting on a night between September 20 and 30, 1888, at 10.30 pm.

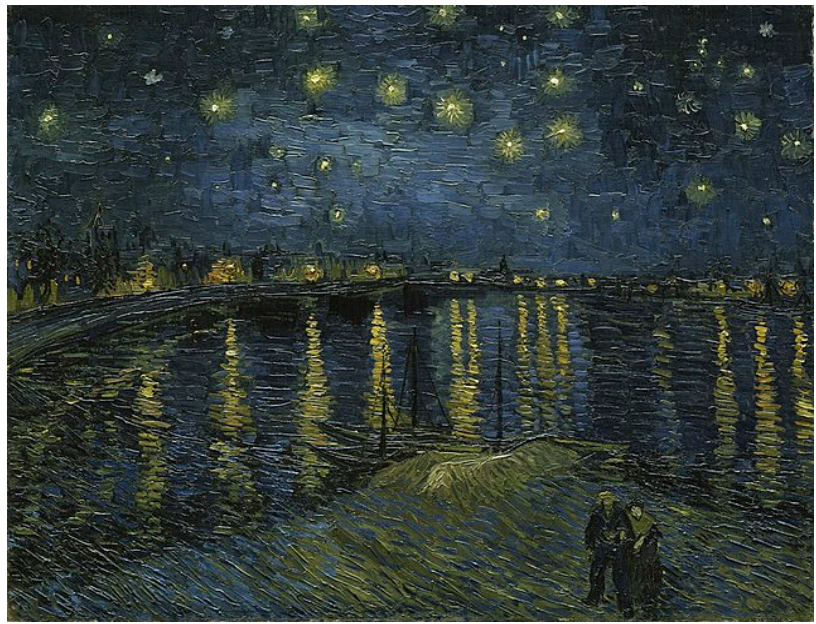

*Figure 2:* **Starry Night over the Rhone** - V. van Gogh, Arles, September 1888 - Musée d'Orsay. The position of the bright stars in the center of the night sky reproduces the big dipper in the Ursa Major constellation. (The image is from Wikimedia Commons in public domain).

---

[5] http://www.bellatrixobservatory.org/cvaai/37/index.html

Added to this detail is also the discovery of a licensing error, that of having depicted the constellation in a different sky area than the real one, specifically in a southwestern direction rather than the north. van Gogh mentioned this paint several times in letters to Theo, to whom he also sent a sketch of the oil painting. It was sent to Paris to be shown in the Independent Artists' Exhibition of 1889 along with a second work, Irises.

Was van Gogh satisfied with this work? The expression he used in his letter to his sister in September 1888 casts some doubt "*...it's clear that to paint a starry sky it's not nearly enough to put white spots on blue-black*" (1888 Letter to Willemien, 678). This implies that he felt this was not his final work for a starry night but only a first try: an artist can do more than simply placing white spots on blue-black to paint a starry night.

Fascinated by the night sky and having been in the lively city of Paris, where he heard the echoes of the discoveries of the nature, diversity and structures of the stars (in magazines, books circles and cafes), clearly he had the will to dare and go further in his mission to communicate the beauty and secrets of a starry night through art. Artists and astronomers, such as Étienne Léopold Trouvelot and William Parsons the 3rd Earl of Rosse, produced illustrations of the sky seen through new, more powerful telescopes. But an artist like van Gogh felt he could do more, he could use the power of his imagination, i.e. arts, to go beyond what eyes can see, and depict feelings as well as perceptions of new astronomical discoveries. In a letter to Theo, he for example defines himself as an arbitrary colorist, and while describing the atmosphere in Arles his thoughts go to the cosmos:

"*Because instead of trying to render exactly what I have before my eyes, I use colour more arbitrarily in order to express myself forcefully....... And still to feel the stars and the infinite, clearly, up there.*" (1888, Letter to Theo, 663).

Stellar colors are also mentioned in a letter to Theo from Les Saintes-Maries-de-la-Mer :

"*The sky, a deep blue, was flecked with clouds of a deeper blue than primary blue, an intense cobalt, and with others that were a lighter blue — like the blue whiteness of milky ways. Against the blue background stars twinkled, bright, greenish, white, light pink — brighter, more glittering, more like precious stones than at home — even in Paris. So it seems fair to talk about opals, emeralds, lapis, rubies, sapphires.*" (1888, Letter to Theo, 619)

Van Gogh lived the scientific climate of the second half of the XIX century after his visit to Paris, and was at the same time fascinated by the relationship between stars and the destiny of our souls. In the Astronomie Populaire it is written:

"*The star light, which glimmers sometime vividly, sometime feebly, in intermittent gleams, sometime white, green or red, like the flashing fires of a limpid diamont.... Spectral analysis applied to the double stars have proved that the beautiful colors are not due to contrast but are real.*" (Flammarion, 2007).

Stellar spectroscopy, the study and classification of spectra, was born early in the 19th century when the German scientist Joseph von Fraunhofer discovered dark lines in the spectrum of the Sun. Following Fraunhofer, other astronomers in the second half of the 19th century, such as Gianbattista Donati and Angelo Secchi SJ in Italy, analyzed stellar spectra and developed classification schemes. Supported by his equipment Secchi was able to publish Le Stelle in 1877 in which he reported the classification of at least 4,000 stars into five groups according to spectral characteristics. He wrote that Type 1, Type 2 and Type 3 stars comprise bluish-white, yellow and reddish-orange stars respectively. In the French edition (Secchi 1880) a figure labelled " Etoiles Colorées" shows several stars with their intrinsic color. Stellar classification schemes as well as the description of the colors and spectra of a few known stars and double stars are reported in the chapter of Flammarion's Popular Astronomy dedicated to the light of the stars:

> "...*instead of being white the stars often shine with colored light, showing in their strange couples admirable association of contrasts where the astonished eye sees the fires of the emerald united with those of the ruby, of the topaz with those of the sapphire, of the diamond with the turquois, or the opal with the amethyst , thus sparkling with all the tints of the rainbow.*" (Flammarion, 2007).

The concept that our sun is similar to other stars in the heavens and therefore that stars have a structure and might change brightness during their life is also well underlined in Flammarion's book.

The first reference to the famous Starry Night is in an early June 1889 letter to Theo where Vincent writes:

> "*This morning I saw the countryside from my window a long time before sunrise, with nothing but the morning star, which looked very big........To not be indifferent and not exhibit something too mad, perhaps the* starry night *and the landscape with yellow greenery which was in the walnut frame.*" (1889, Letter to Theo, 777).

In The Starry Night (see Figure 3) the stars are depicted not merely as little yellow dots against a dark sky, they exhibit a structure and become non-uniform in color going from their center to the edges. Colors change from one star to another. The planet Venus, described by van Gogh as ''the morning star'', was the most luminous object in the sky at the time this masterpiece was created. Positioned to the right of the cypress tree, Venus is portrayed with a prominent white envelope surrounding a small yellowish core. According to STELLARIUM, the planet, with its apparent magnitude of -4.6 reached its peak luminosity during June 1889. The starry night is reported just before dawn, looking towards East in Saint Remy de Provence; this is what the artist could see from his window. In upper side of the canvas, just above Venus and the swirl, three stars might represent the constellation Aries. Aries, the first sign of the zodiac, is van Gogh's own zodiac constellation. The name ''Aries'' meaning lamb, carries a secondary connotation of the Lamb of God symbolizing eternal life. These associations might have further inspired van Gogh in creating The Starry Night. The moon is waning crescent, and its phase, together with the position of Venus, suggests that the painted night sky reflects the view from van Gogh's window close to the 1889 summer solstice just before dawn (see the simulated sky with the help of STELLARIUM). Art historians have long agreed that Vincent van Gogh painted The Starry Night between June 16 and 18, 1889, in Saint-Rémy-de-Provence in southern France. However the moon's phase is indicative of a later night, possibly around June 21, 1889 (approximately 3:00-3:30 am CET or 41 min earlier according to van Gogh's local time). Evidence for dating van Gogh's works primarily stems from his correspondence with relatives and colleagues. Nonetheless, it is plausible that the painting work spanned multiple nights, thus extending the execution of The Starry Night across the summer solstice.

Halos indicate eternity. Stellar halos in The Starry Night are broken concentric circles which surround the ten stars and the planet Venus. They may symbolize eternity but also serve to mimic different luminosities of celestial objects. We underline again that this echoes Flammarion's illustrations, where stars, like the sun, have structures and atmospheres: stars are no longer shining dots in a dark sky. The resonance with stellar classification, which highlighted intrinsic differences among stars, is reflected in this masterpiece. The artist employed diverse palettes to emphasize the varied properties of each star, illustrating their unique characteristics through a harmonious blend of colors.

The scintillation of stars is also described in Flammarion's Popular Astronomy as a phenomenon that depends on the properties of the stars as well as on the Earth atmosphere. Just before introducing stellar spectra it is stated:

> "...*the light that glimmers sometime vividly sometime feebly in intermittent gleams, sometime white, green or red, like the flashing fire of a limpid diamond, seem to animate the interstellar solitudes...we divine better the distant life which is in motion round each of these brilliant fires burning in Infinitude.*" (Flammarion 2007).

We cannot exclude that observing stars close to the horizon van Gogh has experienced the twinkling of stars with their change of colors described in the Popular Astronomy book.

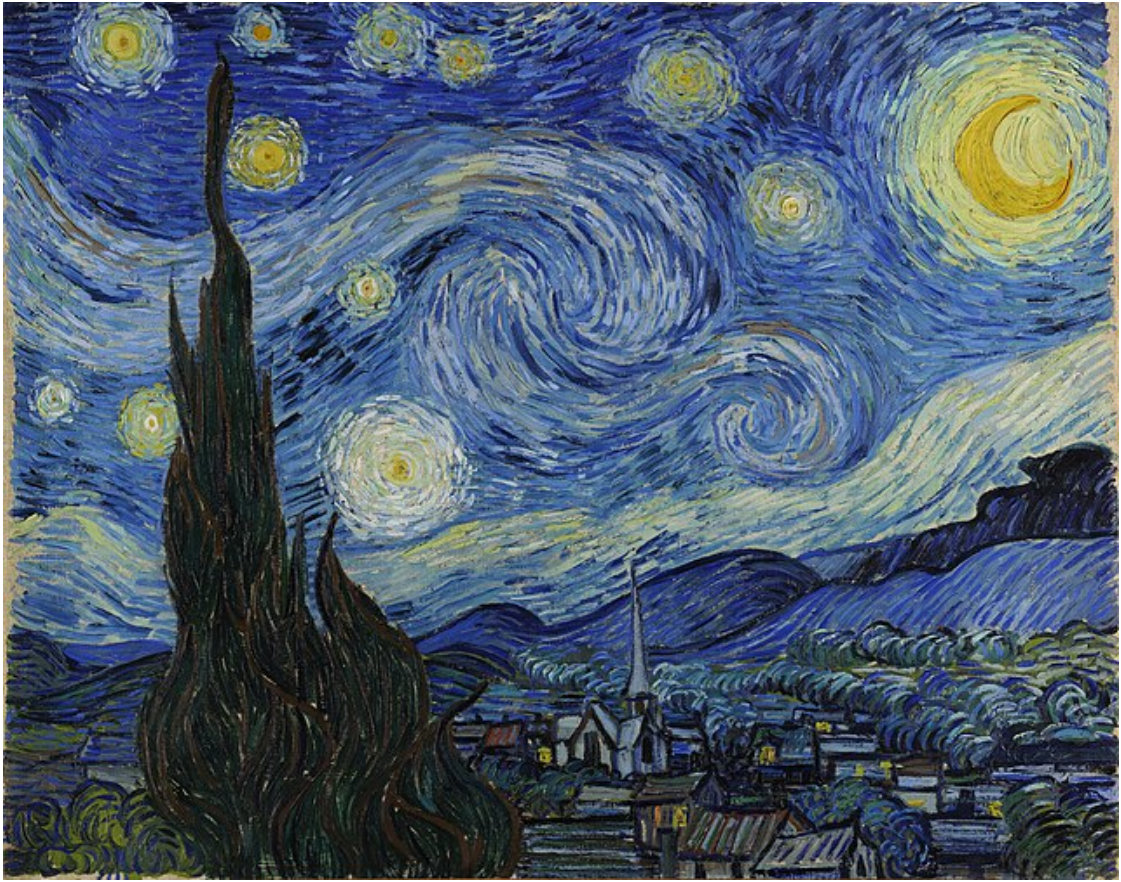

*Figure 3 - Caption:* **The Starry Night** – Oil on canvas by Vincent van Gogh, Saint-Rémy-de-Provence June 1889 - Museum of Modern Art, New York. It depicts the view from the east-facing window of his asylum room just before sunrise, with the addition of an imaginary village and a large swirl in the sky. (The image is from Wikimedia Commons in the public domain).

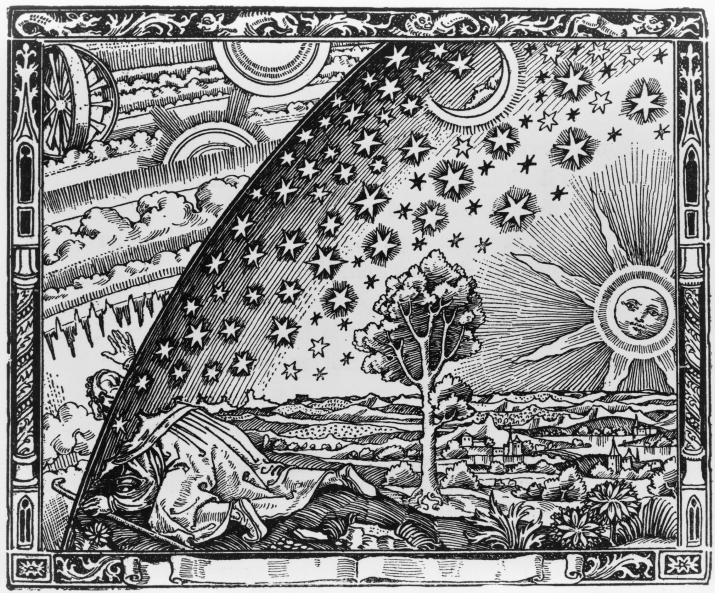

*Figure 4 - Caption:*
The Flammarion engraving is a wood engraving by an unknown artist, reported in the book *L'atmosphère: météorologie populaire*, by C. Flammarion (1888). The missionary is leaving behind a sky where stars are fixed and eternal and is entering a new sky no longer static but in movement, as indicated by the wheels, with the sun and the stars exhibiting an internal structure.
(The image is from Wikimedia Commons in the public domain).

## 5. The dynamic and swirling skies

In its latest work on the starry night (The Starry Night, Figure 3) van Gogh represents the sky with swirling movements. The swirling, and more generally turbulent shapes that characterize some of the van Gogh's latest landscapes and skies, have often been interpreted as symptoms of the psychological distress the painter experienced. However, astronomical reconstructions reveal that the sky visible from his window was quite similar to the one depicted on the canvas, except for the swirl, proving some realism in the painting, Looking at other masterpieces, we see that van Gogh's brushstrokes often take us on a journey around the painting as they twist, turn, and swirl around the subject. This gives a sense of movement and vibration. This technique makes the sky dynamic, reminiscent of the new sky in Flammarion's engraving. The concept of the sky as a dynamic entity was a widespread notion in Paris during the years van Gogh resided there. The dynamic sky is also present in other van Gogh works from the 1889-90 period, such as Cypresses, Road with Cypress and Star, Olive Trees, Wheat Field with Cypresses.

Van Gogh transfigured reality into geometric forms, using a "fluid" geometry, which becomes more evident by the comparison between his paintings and the images produced by recent scientific visualization technologies. Recent papers, for example by Ma et al. (2024) and Wright (2019), analyzing the swirls in The Starry Night found turbulent properties matching the physical law that governs turbulent flows, as observed for example in molecular clouds that give birth to stars. While this might suggest the artist's careful observation of real flows, it does not explain the reason for placing a large swirl at the center of this canvas. Van Gogh's art has always been particularly attentive to the metamorphoses of nature, but it also leaves freedom to communicate feelings and concepts. The large swirl at the center of the starry night might then have no real counterpart but just be part of the artist's attempt to represent a dynamic sky i.e. a sky that evolves, filled with movements and colors. Swirls might represent van Gogh's understanding of the cosmos as a living, dynamic place.

Boime (1984) interprets the swirling figure in the central portion of the sky in *The Starry Night* to represent either a spiral galaxy or a comet, since photographs of these types of objects have also been published in popular media. He asserts that the only non-realistic elements of the painting are the village and the swirls in the sky. Boime notes an important difference concerning the phase of the Moon: on June 19 it was between the half moon and the third quarter, thus resembling more a rugby ball than the fine crescent represented by van Gogh. However we argue that this might just mean that The Starry Night took several nights to be completed, being the moon's phase on June 21, 1889 similar to what the artist paints (see STELLARIUM). Boime underlines that the painter is in closer relationship with the history of science, and suggests van Gogh interest in astronomy, grew through the reading of illustrated popular works - in particular those of Camille Flammarion. The US artist Benson agrees that this interest led him to be the first artist to depict a spiral galaxy:

*"He was quite simply passionate about celestial things. An avid reader of the popular magazine L'Astronomie published by Camille Flammarion, he had discovered with wonder the first photographs of spiral nebulae – those " blue-white milky ways" he talks about in his correspondence. We also know that he had read the Astronomie Populaire of 1881, and the study on Venus that Flammarion had published in March 1889 in the journal l'Illustration."* (Benson 2014).

The Harvard astronomer Whitney conducted his astronomical study of *The Starry Night* independent of Boime (who spent almost his entire career at U.C.L.A.). While Whitney does not share Boime's certainty about the constellation Aries, he concurs with Boime on the visibility of Venus in Provence at the time the painting was executed. He also sees the depiction of a spiral galaxy in the sky, giving credit for the original depiction to William Parsons, 3rd Earl of Rosse. Whitney also theorizes that the swirls in the sky could represent wind, evoking the mistral that had such a profound effect on van Gogh during the twenty-seven-month period he spent in Provence.

By looking at The Starry Night one may wonder why the artist locates the swirl in that part of the sky. Is this just a place at the center of the canvas to give relevance to a subject that he represents with a swirl? Or there is something real in that part of the sky that the artist represented with a large swirl or that hints at a spiral nebula?

The interpretation of the swirl representing a spiral nebula might be valid because van Gogh might have been aware of the scientific discussions around the nature and distances of spiral nebulae. The large swirl in The Starry Night brings to mind motion, as it is also underlined by Nasim (2009) in relation to the engraving of M51 by Lord Rosse (1850), one gets the impression of movement, which however was too difficult to measure in the mid-nineteenth century. Nasim (2009) reports that although many astronomical books and papers in the mid-nineteenth century connect the spiral shape to dynamical models, there have been different speculations favoring either the non-stellar nature and nearby distance of spiral nebulae (like converging cometary objects) or their connection with stellar systems and/or the formation of stars and planetary systems. The subsequent discovery of Huggins and Miller (1864) of emission-line spectra for planetary nebulae, showed that nebulae could be divided into two distinct classes: those that were clouds of luminous gas showing emission lines, and those that were stellar aggregations having only absorption lines. Huggins and Miller (1864) acquired a spectrum of M31 finding no indications of bright lines, although the spectrum truncated in the orange and red parts due to the low luminosity of the nebula. At that time M31 was not known to be a spiral nebula and no other spectra of spiral nebulae have been reported.

The discovery of M31's spiral structure was made by Roberts (1888) who succeeded in photographing its faint disk. The use of photographic plates after 1880 confirmed the spiral character of some nebula, leading to the publication of various theories on their formation (e.g. 1889 referring to the formation of a vortex through the collision of streams of meteorites after observing M81). However, obtaining spectra of spiral nebulae remained difficult due to their faintness. At the time van Gogh visited Paris astronomers were still wondering whether these were gaseous or stellar in nature. Another relevant question concerned the size of the universe: were these systems located within the Milky Way or were they ''island universes'' much further out, as Wright (1750) suggested in his book An Original Theory or New Hypothesis of the Universe? The spiral nebula M51, as drawn by Lord Rosse, is not shown in Flammarion's book Astronomie Populaire, but the concept of clusters and nebulae floating in the depths of the sky, in a universe where nothing is fixed, is there. The M51 engraving is reported in other popular books such as in Arago's book (1854) and in Secchi's book (Secchi 1877). Flammarion reproduces a similar sketch of M51 in an earlier book (Flammarion 1867). Van Gogh might have drawn a generic object with a spiral shape, much more extended than individual stars, inspired by the M51 sketches. He might have placed the swirl at the center of the painting to give more relevance to this class of nebulae. The swirl in The Starry Night winds in the opposite direction to the M51 image, which could be because van Gogh's intention was not to reproduce the M51 sketch exactly or because he did not have the historical drawing at hand. The absence of colors of the swirl may symbolize a lack of information or understanding, reflecting the ongoing mystery the true nature of these celestial objects.

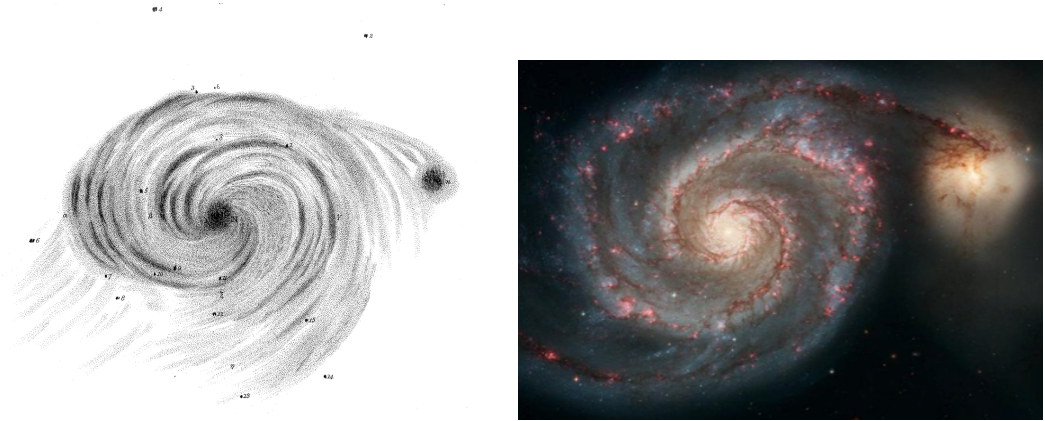

*Figure 5 - Caption:* Left: the engraving sketch of the Great Spiral (M51 or Whirlpool galaxy) by The Earl of Rosse in 1850. Right: the M51 optical+Hα composite image taken in recent years with the Hubble Space Telescope. (Left: The image is from Wikimedia Commons in the public domain- Right : NAS, ESA, and the Hubble Heritage team).

In the region of the sky where van Gogh places the swirl, there are two large spiral nebulae, M31 and M33, our closest spiral galaxies. Due to their proximity these spiral galaxies, especially M31, are more extended than the moon although only a tiny portion of them can be seen by naked eyes. van Gogh might have been aware of M31's location in that part of the sky and might have placed the large swirl there to represent the Andromeda Nebula. Due to its angular extend and low luminosity, M33 needed extreme sky conditions to be observed clearly. Its spiral shape was only definitively revealed in 1895, thanks to advances in photography (Roberts 1895).

On the other hand, the largest object in the sky is the Milky Way, although its spiral shape is not visible to us because we are located within it. Its disk lies in projection about 20° east of the Aries constellation. Could the swirl in van Gogh's painting indicate his intuition or knowledge about the possible Milky Way's possible spiral structure? In 1852, Stephen Alexander (Alexander 1852) proposed that the Milky Way might have a spiral form. The artist, being an avid reader and having worked in a bookstore, may have encountered such hypotheses in his reading.

In Flammarion's Popular Astronomy book is present the concept that some nebulae might represent agglomerates of billions of stars, or else they could consist of cosmic matter that might form new stars, thus connecting present to the past. These concepts, outlined with a great sense of mystery and contemplation, can be found in scientific papers published in the latter half of the XIX century. For example in his paper *Photograph of the Spiral Nebula M33 Trianguli*, Isaac Roberts (Roberts 1895) concludes that the spiral structure of M33 likely resulted from a collision between two swarms of meteorites moving in opposite directions, a hypothesis already suggested a few years earlier for M81 (Roberts 1889). Meteorites and motion were considered to be responsible for the spiral shape.

The position of the swirl in The Starry Night coincides intriguingly aligns with the location in the sky of the radiant point of a meteor swarm called the Arietids. This swarm, active from mid-May to late June, is a significant meteor shower comparable in activity and duration to Perseids and Geminids though less know due to its peak occurring in the very early morning hours. In a recent paper, Abedin et al. (2017) conducted numerical modeling of the Arietid stream to identify the parent body and to constraint its age. They concluded that the most plausible scenario for the formation of the Arietids is a continuous cometary activity of 96P/Machholdz over an interval of 12,000 years. Therefore, the meteor shower would have been visible during van Gogh's observations. Historical comets are described in Flammarion's books, and sometime they were announced as disasters for the fate of Earth, while astronomers were competing for discovering new ones. The connection between comets and shooting stars can be found in Astronomie Populaire. The earlier suggestion that the swirl might represent a comet gains significance due to the presence of the Arietids in that part of the sky.

To summarize, the eight hypotheses we discussed regarding the possible significance of the swirl in The Starry Night are the following:
   a) dynamic sky,
   b) the spiral galaxy M51,
   c) a comet,
   d) the mistral wind,
   e) a generic and arbitrary spiral nebula,
   f) one of the extended nearby extragalactic nebula located in that part of the sky,
   g) the Milky Way,
   h) the Arietids meteor shower.

While these hypotheses provide strong arguments that the swirl is not a product of van Gogh's psychological distress, but of his cultural background, attention and fascination for the night sky, with the artist's imagination allowed to go beyond what eyes can see, it remains challenging to draw a definitive conclusion. I am inclined to favor hypotheses a) + e) + h). The artist must have noticed the meter shower in that area of the sky and during that time of the year. At the same time he was likely aware of the new astronomical concepts, such as the dynamic nature of the discovery of spiral nebulae. The shooting stars could be seen as a symbol of this dynamic sky, reflected also in the spiral nebulae. These nebulae were linked to cosmic collisions during the late nineteenth century, such as those between swarms of meteorites and the formation of new planetary systems. Such celestial objects were captivating and paved the way for the discovery of extragalactic space.

**6. Celestial conjunctions**

Van Gogh's fidelity in reproducing the positions and characteristics of celestial objects as well as his keen attention to celestial phenomena can be seen in several of his artworks, particularly during the last years of his life. For istance, Olson and Doescher (2001) calculated that van Gogh painted *White House at Night* on June 16, 1890, based on the position of the "star" depicted in the painting (see Figure 6). Their research suggested that this ''star'' must be Venus, which was very bright in the evening sky during June 1890.

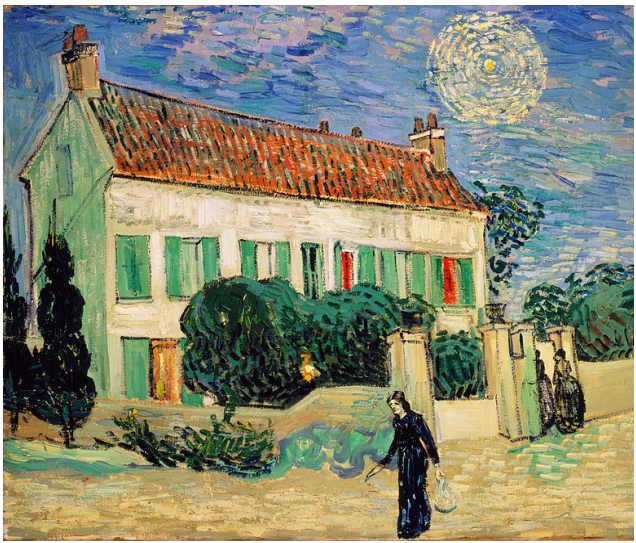

*Figure 6 - Caption*: **White house at Night** is an oil-on-canvas painting created on June 16 1890 in the small town of Auvers-sur-Oise by Vincent van Gogh, six weeks before his death. It is displayed at the Hermitage Museum of St. Petersburg. (The image is from Wikimedia Commons in the public domain)

Astronomical phenomena, like planet conjunctions, are also present in van Gogh's masterpieces. On June 17 1890, just one day after the dating of the White house at Night, van Gogh wrote to Paul Gauguin from Auvers-sur-Oise to describe his Road with Cypress and Star work (see Figure 6):

"*A night sky with a moon without radiance, the slender crescent barely emerging from the opaque shadow cast by the earth. One star with an exaggerated brilliance, if you like, a soft brilliance of pink and green in the ultramarine sky, across which some clouds are hurrying* ". ((1890, Letter to Paul Gauguin, RM23)

There was no lunar eclipse during that period and van Gogh appears to have been confused between the phases of the moon and the phenomenon of a lunar eclipse, as he mentioned in the letter to Gauguin. However, using the astronomical software STELLARIUM, we discovered that on April 23, 1890, there was a conjunction between Venus and Mercury, with the planets separated by only 2º in the sky. If the dim shining object at the bottom left of Venus represents Mercury, and the moon was crescent and slightly higher than Venus, similar to what is depicted in the painting shown in Figure 7, the date of the work could be April 20, 1890, around 8 pm CET (7:19 pm local time), a few days before the conjunction. At that time the moon had a very thin illuminated surface visible from the Earth, positioned at about 4º to the left of Venus, while Mercury was lower to the right of Venus. Interestingly, van Gogh's depiction of these three celestial objects appear to be mirrored compared to the STELLARIUM simulation (except for the moon's illuminated part). Art historians (e.g. Erickson 1998) suggest that this masterpiece was created in May 1890. It is therefore likely that van Gogh mentally framed the planetary conjunction and later attempted to reproduce the view, albeit with some artistic license.

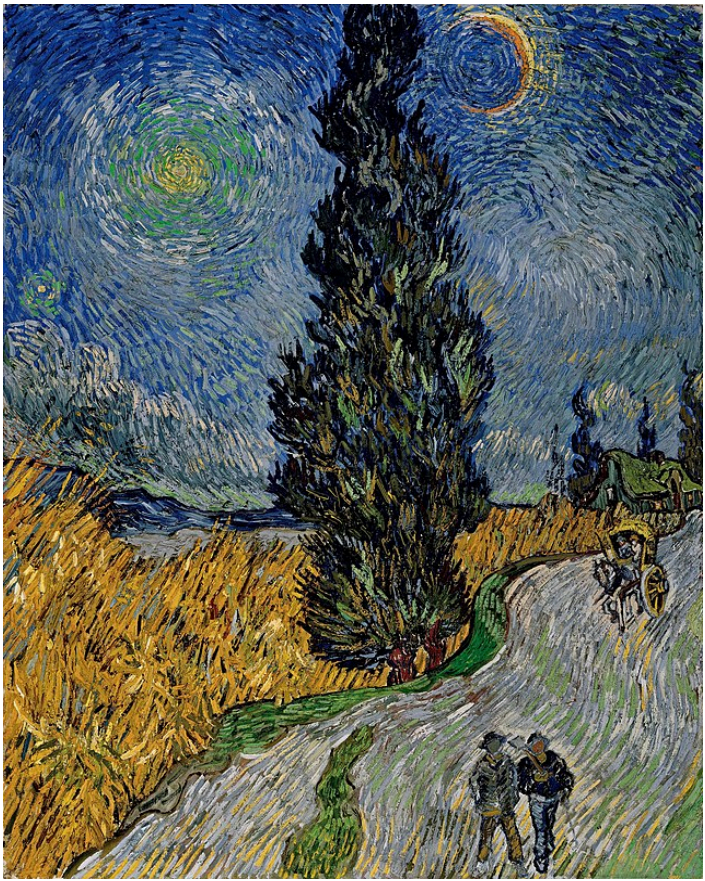

*Figure 7 - Caption*: **Road with Cypress and Star** by V. van Gogh resides in the Kröller-Müller Museum and it is dated May 1890. (The image is from Wikimedia Commons in the public domain)

And finally we would like to emphasize another planetary conjunction possibly present in what has often been claimed to be van Gogh's last work: Wheatfield with Crows created in Auvers-sur-Oise (see Figure 8).
Evidence from van Gogh's letters suggests that Wheatfield with Crows was completed around July 10, 1890. There have been claims that he may have finished other works, such as Tree Roots, even later, so it remains uncertain if this was truly his last painting. The artwork features a dramatic, cloudy sky filled with crows soaring over a wheat field, with a central path leading to nowhere. The flight of the crows merges with a dramatic blue and black sky, hinting that it is getting dark. Notably, two round whitish objects are clearly visible at the horizon.  Around July 10, 1890, van Gogh wrote to his brother Theo stating that he had painted three large canvases since visited Paris on July 6. He described two of these as immense stretches of wheat fields under turbulent skies, emphasizing a sense of sadness, and later mentioning "extreme loneliness". One of these could potentially be Wheatfield with Crows.

The larger whitish object towards the end of the path may represent a setting sun or a luminous star at the end of the road, which is where the road is going. This idea resonates with what he wrote (see Section 2): "....*Just as we take the train to get to Tarascon or Rouen, we take death to reach a star*.....". The second whitish object might have been added later, as some of white/light blue paint spills over one of the crows. In July 1890 a spectacular planetary conjunction between Saturn and Venus was visible with the two planets were separated by less than 15 arcminutes (approximately half the size of the moon) on July 17, 1890.  Therefore, the object to the left of the setting sun  could represent these planets. According to STELLARIUM, on that evening, an extremely thin, barely visible crescent moon, was setting very close to the horizon where the sun set. The moon, Saturn and Venus were all setting around the same time on July 19, 1890. Although we cannot exclude the possibility that the largest object is a crescent moon, the roundness of the object makes this hypothesis more unlikely. The western horizon was very noteworthy during mid-July 1890. Van Gogh might have mentioned the canvas before completing it or  might have  added the second celestial object near the horizon at a later time. The presence of two whitish celestial objects close to the horizon - one  remarkably positioned  at the end of the path - highlights the artist's keen attention to celestial phenomena and his deep connection to the sky as a physical, mystical and religious entity.

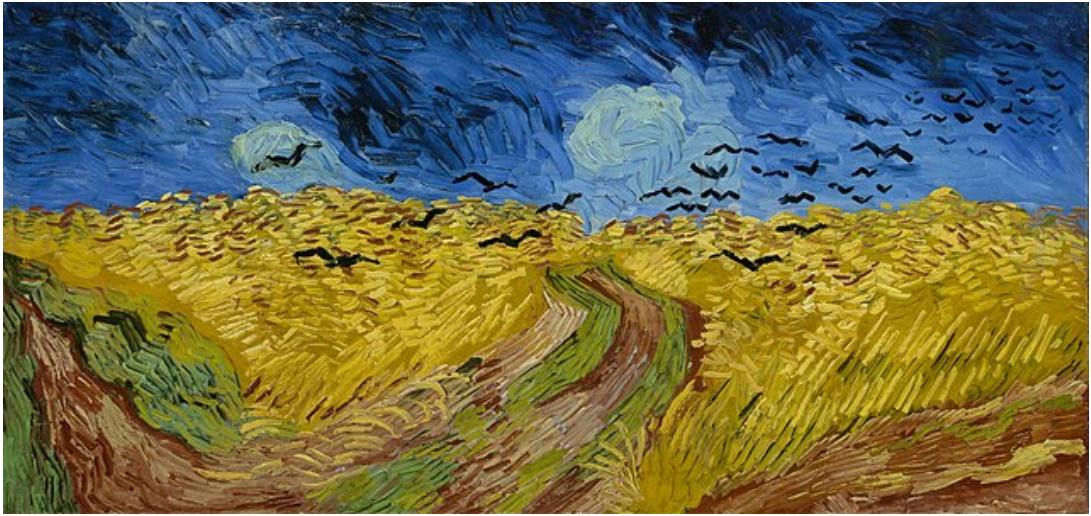

*Figure 8 - Caption*: **Wheatfield with Crows** - oil on canvas by Vincent van Gogh, Van Gogh Museum, Amsterdam (The image is from Wikimedia Commons in the public domain)

**7. Summary and conclusions**

In this paper we have shown van Gogh's ongoing interest to astrophysical discoveries and the deep fascination he felt for celestial phenomena, often depicted in his masterpieces. Previous analyses have already underlined the presence of elements in van Gogh's art linked to the developments of astronomy in the second half of XIX century, or the correspondence of the positions of celestial objects with what he observed at night. We have added in this paper new elements, such as the 1889 lunar eclipse visible in southern France in the Moonrise painting, and planetary conjunctions. Other elements have been related to the period he spent in Paris, such as his awareness of stellar diversity through spectroscopy and classification, and the astronomers' wonder about the nature of nebular objects. By analyzing the swirling skies in his famous Starry Night we speculated on additional new possible significance, real correspondences and source of inspirations for van Gogh's most popular masterpiece.

***Acknowledgements****:* In this work we have used the STELLARIUM Astronomy Software. The author would like to thank Bruce Elmegreen for having suggested the meeting at the Museum of Modern Art New York in 2008 to discuss a referee report. The museum's extemporaneous exhibit on van Gogh has been inspiring for the work presented in this paper. Bruce Elmegreen and Simone Bianchi have the author's gratitude for their encouragement to publish this work and for their comments to a first draft of this manuscript.